# A Comprehensive Model of the Degradation of Organic Light-Emitting Diodes and Application for Efficient Stable Blue Phosphorescent Devices with Reduced Influence of Polarons

Bomi Sim[1], Jong Soo Kim[3], Hyejin Bae[3], Sungho Nam[3], Eunsuk Kwon[3], Ji Whan Kim[3], Hwa-Young Cho[3], Sunghan Kim[3]*, Jang-Joo Kim[1,2]*

[1]*Department of Materials Science and Engineering, Seoul National University, Seoul 151-744, South Korea.*

[2]*Research Institute of Advanced Materials (RIAM), Seoul National University, Seoul 151-744, South Korea.*

[3]*Samsung Advanced Institute of Technology, SEC, 130 Samsung-ro, Yeongtong-gu, Suwon-si, Gyeonggi-do, 443-803, Korea.*

* E-mail: jjkim@snu.ac.kr, shan0819.kim@samsung.com

**Abstract**

We present a comprehensive model to analyze, quantitatively, and predict the process of degradation of organic light-emitting diodes (OLEDs) considering all possible degradation mechanisms, i.e., polaron, exciton, exciton–polaron interactions, exciton–exciton interactions, and a newly proposed impurity effect. The loss of efficiency during degradation is presented as a function of quencher density, the density and generation mechanisms of which were extracted using a voltage rise model. The comprehensive model was applied to stable blue phosphorescent OLEDs (PhOLEDs), and the results showed that the model described the voltage rise and external quantum efficiency (EQE) loss very well, and that the quenchers in emitting layer (EML) were mainly generated by dopant polarons. Quencher formation was confirmed from a mass spectrometry. The polaron density per dopant molecule in EML was reduced by controlling the emitter doping ratio, resulting in the highest reported $LT_{50}$ of 431 hours at an initial brightness of 500 cd/m$^2$ with $CIE_y$<0.25 and high external quantum efficiency (EQE) >18%.

## I. INTRODUCTION

Organic light-emitting diodes (OLEDs) are widely used as displays in cellular phones and TVs and their application is extending to lighting. The efficiency of blue OLEDs in these devices is still low compared to phosphorescent red and green OLEDs[1–4] and the device lifetimes of triplet harvesting phosphorescent and TADF (thermally activated delayed fluorescent) blue OLEDs are too short to be used in real displays.[5–11] The development of highly efficient pure blue OLEDs with long lifetimes is a key research topic. Various mechanisms, such as exciton–exciton[8] or exciton–polaron interaction,[6,12,13] exciton itself,[13–15] and radical ion pair between host exciton and dopant[9] have been proposed to explain the degradation of (blue) OLEDs. A few degradation models of OLEDs have also been proposed. Giebink et al. reported a model of luminance loss and voltage rise in blue PhOLEDs and proposed exciton–polaron annihilation as the primary mechanism of defect formation.[6] Brutting et al. and Homles' group presented luminance loss models that depict the relative ratio of exciton formation efficiency and effective quantum efficiency on degradation of green PhOLEDs.[16–18] A model describing the degradation processes in polymer LEDs has also been reported.[19]

As quencher formation is likely to be caused by various factors simultaneously, it is necessary to consider a combination of the previously proposed mechanisms. To develop a general degradation model, it is necessary to include additional factors that may contribute to quencher formation, such as unstable polaron states without involving excitons, and impurities incorporated during device fabrication ($O_2$, $H_2O$ or other volatile components) and/or in source materials. Despite the potential instability and high density of polaron state in EML during operation, the presence of polarons in EML has not been reported as a major reason for degradation. In addition, rapid changes in driving voltage and luminance over a short time scale can be caused by impurities, such as water and oxygen.[5,7] However, conventional degradation models did not consider the impurity factors. Therefore, a model that can describe the quantitative contribution of each mechanism to degradation and explain the more general and comprehensive degradation phenomena, including quencher formation, is still needed.

Here, we present a comprehensive and general model for describing the degradation of OLEDs. Our model consists of two equations describing the efficiency loss and the voltage increase during electrical operation as functions of quencher density. The effect of the quencher on exciton formation efficiency and the effective quantum efficiency are described by measurable parameters in the equations. Polaron and impurity effects as well as the exciton, exciton–polaron, and exciton–exciton interactions are considered to be the mechanisms underlying quencher generation (degradation). This model allows quantitative analysis of the contribution of each mechanism to the total device degradation. Analysis of a stable blue phosphorescent device using the degradation model showed that quenchers are generated by all mechanisms originating from impurities, polaron, exciton, exciton–polaron, and exciton–exciton interactions. Transporting materials are degraded by the exciton-mediated processes. Interestingly, however, quenchers in the emitting layer (EML) are mainly generated by dopant polarons. To our knowledge, this is the first report that the polaron itself can have the greatest impact on degradation of phosphorescent dyes in EML. In contrast, transporting materials are degraded by the exciton-mediated processes. We then reduced the polaron density per dopant molecule by increasing the emitter doping concentration to achieve the highest reported lifetime of 431 hours of $LT_{50}$ at an initial brightness of 500 cd/m$^2$ while maintaining the high EQE around 18% with CIEy<0.25 in a blue phosphorescent device.

## II. MODEL

### A. Modeling of degradation process

The processes of exciton generation and annihilation in an OLED are schematically illustrated in Fig. 1. The sky blue region represents the EML, while yellow and orange colors represent the other layers. The processes in the pristine device are indicated by the blue lines. The injected charges are recombined at host (H) (①) or at dopant (D) (③) with recombination rates of $k_{rec}^{H}$ and $k_{rec}^{D}$, respectively, or at other layers or leaked out from the EML (⑨). The exciton energy of the host is transferred to the dopant at a rate of $k_{ET}^{HD}$ (②) or emitted as light or heat with decay rates of $k_{r}^{H}$ and $k_{nr}^{H}$ (⑤), respectively. The dopant excitons generated by direct recombination on the dopant or energy transfer from host, decay radiatively or non-radiatively with decay rates of $k_{r}^{D}$ and $k_{nr}^{D}$, respectively (④).

The newly developed processes in the aged (degraded) device are indicated by red lines. Quenchers indicated by Q in Fig. 1 are assumed to be generated throughout the electrically degraded device, including the EML. We assume that the singlet-triplet ratio (1:3), radiative decay rates, energy transfer rate from host exciton to dopant, absorption coefficient in EML, refractive indices, and Purcell factor are constant during the electrical operation of OLEDs. The quencher (Q) can act as deep charge trap, non-radiative recombination center, and luminance quencher depending on their energy and location in the device. The rates of luminance loss processes are affected by the density of the quenchers in the EML. The host excitons can transfer energy to the quencher (⑥) with a rate of $k_{ET}^{HQ}[Q(t)]_{EML}$, which reduces the energy transfer efficiency to the dopant ($\eta_{ET}^{HD}$). Here, $[Q(t)]_{EML}$ is the quencher density in EML. The energy transfer from the dopant exciton to the quencher (⑦) with a rate of $k_{ET}^{DQ}[Q(t)]_{EML}$, and biparticle interaction (⑩) with an annihilation rate of $k_{BQ}^{D}$ result in luminance quenching as well as quencher generation. When quenchers act as non-radiative recombination centers, the injected charges can be recombined directly at the quenchers (⑧) with a recombination rate of $k_{rec}^{Q}$ reducing the exciton formation efficiency in the EML. Note that the bimolecular interactions[20–23] and charge leakage[22,23] can also occur in the pristine device.

### B. Modeling of efficiency loss

The EQE of an electrically aged device shown in Fig. 1 can be expressed as follows:

$$EQE(t) = \left(\frac{J_{Cn}(t) - J_{An}(t)}{J}\right) \cdot \left(\frac{k_{rec}^{H} + k_{rec}^{D}}{k_{rec}^{H} + k_{rec}^{D} + \frac{e}{\varepsilon\varepsilon_0} f \mu_n n [Q(t)]_{EML}}\right) \cdot \left(\alpha_{EL}(t)\left(\frac{k_{ET}^{HD}}{k_{ET}^{HD} + Fk_r^H + k_{nr,int}^H + k_{ET}^{HQ}[Q(t)]_{EML}}\right) + (1 - \alpha_{EL}(t))\right) \cdot$$

$$\left(\frac{Fk_r^D}{Fk_r^D + k_{nr,int}^D + k_{ET}^{DQ}[Q(t)]_{EML} + k_{BQ}(t)}\right) \cdot \eta_{out}(t) \qquad (1)$$

where $J_{Cn}(t)$ and $J_{An}(t)$ represent the electron current densities at the cathode and anode sides of the EML at operation time $t$, respectively; $\alpha_{EL}$ is the ratio of host excitons to the total excitons (host exciton plus dopant excitons) formed in the EML, and $F$ is the Purcell factor describing enhancement of the spontaneous emission rate in the device.[25] The first and second terms on the right hand side represent the exciton formation efficiency ($\eta_{EF}(t)$), considering recombination at quenchers (details given in Table I). We considered hole traps in the recombination at quenchers, but electron traps can be included depending on devices. The third and fourth terms represent the effective quantum efficiency ($q_{eff}(t)$), which is decreased due to the increased exciton quenching. The general model in Equation (1) describes the effects of $\eta_{EF}(t)$ and $q_{eff}(t)$ on the total EQE loss as a function of $[Q(t)]_{EML}$. We applied the model to analyze the degradation process of a blue phosphorescent device and to enhance further the operational lifetime. All of the rates in Fig. 1 and Equation 1 can be determined experimentally, as summarized in Table I. The equation (1) can be applied to phosphorescent OLEDs as it is, but it can also be applied to fluorescent [need to multiply by 0.25 to the EQE(t)] or TADF OLEDs with slight modification.

### C. Modeling of voltage rise and quencher density

The changes in operating voltage during electrical operation arise from trapped charges generated both inside and outside of the EML. Assuming that all the quenchers ($[Q(x,t)]$) act as deep charge traps, the voltage rise is represented as[6,8]

$$\Delta V(t) = \frac{e}{\varepsilon \varepsilon_0} f[Q(t)] \int_0^L x \cdot g(x) dx . \qquad (1)$$

Here, $\varepsilon, \varepsilon_0$ are the relative dielectric constant of the organic layers and permittivity of free space, respectively, $f$ is the occupational probability of charge at Q, and L is the thickness of OLEDs, respectively. The quencher density at positon x and time t has a relationship of $[Q(x,t)] = [Q(t)] \cdot g(x)$, where g(x) is the normalized quencher distribution function as shown in Appendix A in details. As quenchers can be formed by polarons, excitons, exciton–polaron interaction, and exciton–exciton interaction, it is necessary to consider the contribution of all possible mechanisms. In addition, we consider the impurity effect (or more generally extrinsic effect) with the initial concentration of $[A(x)]_0$ in the degradation model, which originates from source materials or incorporated during the fabrication process, including $H_2O$ or $O_2$. This is necessary because the lifetime of OLEDs is significantly influenced by purity of materials and fabrication processes, e.g., vacuum level in the evaporation process or residual solvent in the solution process.[5,25,26] Here, we assumed that these impurities formed quenchers at the initial stage of the operation with first-order kinetics by ( $A + P \xrightarrow{k_{QF}^{imp}} Q$ ), where A is the impurity density, P is polaron density, Q is the quencher density, and $k_{QF}^{imp}$ is the quencher formation rate constant by impurities, resulting in $[Q(x,t)]_E = [A(x)]_0 \left[1 - \exp(-k_{QF}^{imp}[P(x)]t)\right]$. Different kinetic equations can be applied depending on the nature of impurities. This impurity effect accounts for the rapid increase in driving voltage at the initial stage in our specific example discussed later, and then the quencher density by impurities becomes constant if the impurity is consumed. Then, the quencher formation rate at position $x$ and time $t$ is represented as:

$$\frac{d[Q(x,t)]}{dt} = k_{QF}^P[P(x,t)] + \left(k_{QF}^E + k_{QF}^{EP}[P(x,t)]\right)[N(x,t)] + k_{QF}^{EE}[N(x,t)]^2 + k_{QF}^{imp}[P(x,t)][A(x,t)]$$

(2)

where $[P(x,t)]$ is polaron density, $[N(x,t)]$ is exciton density, and $[A(x,t)]$ is the impurity density with the initial density of $[A(x)]_0$. The quencher generation rates by different mechanisms can be determined by analyzing the voltage rise over time by combining Equations (2) and (3). It should be noted that the quencher generation by excitons and exciton–polaron under constant current shows the same first-order reaction kinetics against the exciton density. The details of the application of the equations are presented in Appendix A. The density, location, and generation rate of quenchers can be obtained by the best fit of the voltage rise model with experimental data.

### III. Results and Discussion

#### A. Application to blue phosphorescent device

Fig. 2(a) shows the schematic device structure[9,27] with the energy level diagram of a blue phosphorescent device used for analysis of the degradation mechanism using the degradation model. The blue-emitting Ir-dopant and the wide bandgap host (mCBP-CN) are used in the EML with the emitter doping concentration of 10 wt%, and their chemical structures are shown in Fig. 2(b). The $J$–$V$–$L$ characteristics are shown in Fig. 2(c). The experimentally obtained maximum EQE of the pristine device was 18.3%, as shown in Fig. 2(d), and the CIE coordinate of the emitted light was (0.15, 0.23) at $J$=10 mA/cm$^2$ (Fig. 2(e)). The exciton profiles in the EML were measured experimentally by the sensing-layer method (Fig. 2(f)) to apply the model.[22]

Fig 3(a) (black line) and Fig. 3(b) (black line) show the rise in operating voltage ($\Delta V$) and luminance loss (or EQE loss) over time at a constant current density of $J$=1.65 mA/cm$^2$ corresponding to the initial luminance of 500 cd/m$^2$. It should be noted that the luminance loss is proportional to the efficiency loss because the emission spectra do not change during the operation. LT$_{50}$, corresponding to the operation time when the luminance decreases to 50% of its initial value, approached 238 hours with the voltage rise ($\Delta V$) of 0.34 V.

### B. Analysis of driving voltage and calculation of quencher density

The generation mechanism and density of the quencher in the EML and charge transporting layers were determined using the voltage rise model. In this simulation, we consider five different quencher generation mechanisms: polaron, exciton, and/or exciton–polaron interactions, exciton–exciton interactions, and impurities represented by the rate constants of $k_{QF}^{P}[p], k_{QF}^{E} + k_{QF}^{EP}[p], k_{QF}^{EE}, [A]_0$ and $k_{QF}^{imp}[p]$.

Fig. 3(a) shows the fitting results of the experimental voltage change (black line) using the model (Equations 2 and 3) developed in this study. Details of the fitting process are described in Appendix A and the results are in Table. II. The voltage increase with time cannot be reproduced well just by considering the polaron, exciton, exciton–polaron, and exciton–exciton interactions ($k_{QF}^{P}, k_{QF}^{E} + k_{QF}^{EP}[P]$ and $k_{QF}^{EE}$) without the impurity effect (blue dashed line). In contrast, the experimental data were very well fitted by including the impurity factor (red dotted-dashed line). The impurity effect $\{[A]_0$ and $k_{QF}^{imp}\}$ accounted for the rapid increase in driving voltage at the initial stage, and the defect generation rate was exponentially reduced with time as the impurities were consumed by reaction. To confirm that the fitted line in Fig. 3(a) is the only solution of the model, we examined various fittings by changing the fitting parameters as shown in Fig. S1. Fig. 3(c) shows the contributions of different degradation processes to the increase in driving voltage. The quencher densities at $LT_{50}$ generated by polaron, exciton–polaron, exciton–exciton annihilation, and impurity factor were $3.4 \times 10^{17} \, (cm^{-3})$, $2.0 \times 10^{17} \, (cm^{-3})$, $2.1 \times 10^{17} \, cm^{-3}$, and $2.1 \times 10^{17} \, (cm^{-3})$, respectively. These results clearly showed that not a single mechanism but all of the mechanisms together contributed to the generation of quenchers in the device affecting the driving voltage.

### C. Quencher generation rate and mechanism in EML

The parameters in Fig. 1 determined using the pristine device using the equations in Table I are summarized in Table III and Experimental section. The exciton quenching rates corresponding to ⑥ and ⑦ in Fig. 1 were obtained by measuring the decreased PL intensity of the host and the radiative lifetime of the emitter exciton, respectively, in the degraded devices at $LT_{75}$, $LT_{60}$, and $LT_{50}$, and are shown in Fig. 4(a) and 4(b). (refer to Table I and Experimental section). The measured quenching rates of the host and dopant excitons and, therefore, the quencher density in the EML, increase linearly with operating time if the energy transfer rate are assumed to be constant ($k_{ET}^{DQ}, k_{ET}^{HQ}$). These results indicate that the quenchers in the EML are generated by polarons because the polaron density is kept constant during the operation. The exciton-mediated processes (exciton, exciton–polaron, and exciton–exciton) were discarded as the origins of quencher generation because the exciton density decreases over time in the EML (Appendix B). The exciton mediated processes were effective outside the EML where the quencher generation rate is reduced with operation time. Polaron-induced quencher generation in the EML was supported by the observation that the lifetime acceleration factor[28–30] is close to 1, as shown in Fig. 4(c). EQEs were almost constant from 500 cd/m$^2$ to 3000 cd/m$^2$ in the device and so the initial luminance was linearly proportional to the current (polaron) density. Therefore, LT$_{50}$ (quencher density) was linearly proportional to the current density. If the degradation is induced by exciton–polaron or exciton–exciton mechanisms, $n$ must be >1 and close to 2. If quenchers are generated by excitons, $n$ must be lower than 1.

### D. Quencher generation rate and mechanisms in transporting layers

Among the total quenchers, we considered that the polaron-induced quenchers and 20% of impurity quenchers (30 nm of EML to total 140 nm of device) are generated in the EML as discussed in the previous section. Therefore, Q of $3.8 \times 10^{17}$ cm$^{-3}$ is in the EML among the total Q of $9.5 \times 10^{17}$ cm$^{-3}$ at LT$_{50}$. The quenchers from exciton interactions analyzed in Fig. 3(c) and 80% of the impurity quenchers are then generated outside of the EML. In this device structure, holes are likely to be accumulated at the NPB/TCTA and TCTA/mCBP interfaces due to the energy barriers. Electrons are also likely to be leaked to the hole transporting layer (HTL) as inferred from the low energy barrier and the large exciton density near the HTL (Fig. 2(f)). In addition, the excitons at the EML/electron transporting layer (ETL) interface and ETL can generate quenchers by exciton–polaron (both anions and cations) and exciton–exciton interactions due to the high density of polarons. Therefore, exciton–polaron and exciton–exciton interactions can result in the formation of quenchers in the transporting layers along with a portion of the impurity quenchers, thus influencing the driving voltage but not the luminance.

### E. Prediction and analysis of the efficiency loss of blue PhOLED

With the model, we can predict EQE and luminance as functions of time and quencher density using the rate constants shown in Table III and Fig. 4. We assume that out-coupling efficiency ($\eta_{out}(t)$) does not change during the degradation because the change in exciton distribution in the EML is negligible as deduced from the EL spectra that remained the same during electrical aging (Fig. 2(e)). Fig. 3(b) shows a comparison of the experimental data (black line) with the theoretical predictions based on different degradation mechanisms. The red dashed line, considering the polaron-induced quencher formation in the EML as the degradation mechanism along with the impurity effect, showed excellent agreement with the experimental data compared to the exciton or exciton–polaron (violet) or exciton–exciton (yellow) interactions. Fig. 3(d) shows the EQE loss over time (blue area) with loss of exciton formation efficiency (green area) and effective quantum efficiency (yellow area) calculated using Q(t) and Equation (1). The symbols are experimental values from fresh, $LT_{75}$, $LT_{60}$, and $LT_{50}$ devices. The experimental values (open symbols) matched perfectly with the calculations (black lines). The exciton formation efficiency was decreased from 0.92 to 0.81 (88%) at $LT_{50}$, and the effective quantum efficiency was decreased from 0.84 to 0.49 (58%) at $LT_{50}$ with $[Q]_{EML}$ of $3.8 \times 10^{17} \ (cm^{-3})$.

### F. More stable blue PhOLED

The above analysis clearly showed that the luminance degradation in the device is mediated by polarons. We increased the doping concentration of emitters in the device to 20% to reduce the quencher formation rate in the EML or to increase the device lifetime. Assuming that the trapped charge density is constant in 10% and 20% doped devices, the charge density per dopant molecule is decreased by half in the 20% doped device, with the expectation of doubling the lifetime. The *J-V-L* characteristics, EQE with the maximum of 18.3%), the emission spectrum and the recombination zone of the 20% doped device were almost the same as the 10% doped device as displayed in Figs. 5(a),5(b) and 2(f), respectively, indicating that the charge transport properties in the EML does not change much with doping concentration in the device. As expected, however, the lifetime of the 20% doped device was two time longer with $LT_{50}$ of 431 hours at an initial luminance of 500 cd/m$^2$ as shown in Fig. 5(c). This is the highest reported value of the lifetime for a blue phosphorescent OLED below CIEy<0.25. An analysis of the degradation mechanisms of the 20% doped device is shown in Fig. 6. The quencher density in the EML of the 20% doped device also increases linearly with operating time, indicating that the quenchers in the EML of the 20% doped device are generated by polarons as like the 10% doped device (Fig 6(a)). The variations of $\Delta V$, [Q(t)] and luminance over time of the 20% doped device are almost same as the 10% doped device as shown in Figs. 6(b), 6(c) and 6(d), respectively, but the lifetime is doubled due to the half of the quencher formation rate compared to the rate of the 10% doped device as shown in Table II.

### G. Quencher analysis by DESI-MS imaging

The above analysis showed that degradation takes place not only in the EML but also in the transporting layers. Polarons and impurities played major role in degradation in EML and other exciton, exciton-polaron and exciton-exciton interactions degraded the transporting layers. Desorption electrospray ionization mass spectrometry (DESI-MS) of the blue devices was performed to confirm the degradation of the materials. The experimental details of DESI-MS are described in the Experimental section. Fig. 7(a) shows DESI-MS images of the fresh and aged ($LT_{10}$) devices with 10% Ir-dopant. The densities of the dopant (m/z=1007.2), TCTA (HTL, m/z=741.3), and DBFPO (ETL, m/z=569.1) of the aged devices were decreased, while the densities of the host (mCBP-CN, m/z=510.2) and mCP (m/z=409.2) remained similar after degradation. The new products with m/z=937.3 and 485.2 were observed only in the aged device, as shown in Fig. S3. Fig. 7(b) shows the possible origins of the new products (m/z =485.2), which may be TCTA molecular fragments generated by dissociation of C-N bonds. The new product with m/z=937.3 could be produced by fragmentation of Ir-dopant ligand and a TCTA molecule. There was no direct evidence of an Ir-containing fragment present only in the aged pixel with the unique isotope pattern of Ir. Nevertheless, the mass density normalized by that of fresh pixels between 10% and 20% Ir-doped devices (Fig. S4.) also showed decreases of Ir-dopant and TCTA after operation. Other degradation products with m/z of 369.1, 493.1, 535.2, 585.1, 645.2, 1528.5, 1607.6, 1612.5, and 1624.6 were also found, and the potential molecular structures of these masses are listed in Table S1. Therefore, the results support the analysis that the dopant is degraded in the EML, and the HTL and ETL are also degraded, as mentioned in the previous sections. However, this mass analysis does not give information on the degradation mechanism of each layer.

## IV. Conclusion

We presented a comprehensive model describing the degradation of OLEDs considering all possible degradation mechanisms, i.e., polaron, exciton, exciton–polaron, and exciton–exciton interactions and a newly proposed impurity effect in an equation. The degradation process is correlated with the formation of quenchers in the model. Therefore, the model allows us to identify the origin, density, and location of the quenchers using the rise of the operation voltage, EQE loss, and some independent experiments. Moreover, variation of the exciton formation efficiency and the EQE during degradation could be obtained without any fitting parameters. We applied the model to analyze the degradation process of a highly efficient stable blue phosphorescent OLED, and the results indicated that the model described the degradation processes of operation voltage and EQE very well. The analysis suggested that the quenchers are generated not by a single mechanism but by all of the mechanisms outlined above. Interestingly, however, quenchers in the EML are generated mainly due to degradation of the dopant by polarons and impurities. The analysis indicated that we could increase the lifetime of the device twice by increasing the doping ratio to reduce the polaron density per dopant molecule to half, and achieve high efficiency (18% EQE) with the highest value of $LT_{50}$ of 431 hours at an initial brightness of 500 cd/m$^2$ for blue phosphorescent OLEDs below $CIE_y<0.25$ using conventional device structure and materials.

## V. Experimental Section

*Materials*

All common layers were composed of commercially available materials, and used without any further purification (sublimed grade). mCBP-CN (>99.96%) and Ir-dopant (>99.88%) were synthesized according to the method reported previously[9,33] and purified by sublimation at $10^{-6}$ torr. The purity of materials was determined by high-performance liquid chromatography (HPLC) analysis (Alliance e2695; Waters Corporation, Milford, MA).

*Device fabrication and characterization*

Blue PhOLED devices were fabricated to analyze quantitatively the performance: indium tin oxide (ITO) (150 nm)/HAT-CN (1,4,5,8,9,11-hexaazatriphenylenehexacarbonitrile) (10 nm)/NPB (N,N-di(1-naphthyl)-N,N′-diphenyl-(1,1′-biphenyl)-4,4′-diamine) (50 nm)/TCTA (4,4′,4″-Tris(carbazol-9-yl)triphenylamine) (5 nm)/mCP (1,3-bis(N-carbazolyl)benzene) (5 nm)/mCBP-CN:Ir-dopant (10 wt%, 20 wt%, 30 nm)/DBFPO (2,8-bis(diphenylphosphineoxide)-dibenzofuran) (10 nm)/DBFPO:Liq (lithium quinolinate) (1:1, 30 nm)/Liq (1 nm)/Al (100 nm). The organic, Liq, and metal layers were deposited sequentially on pre-cleaned ITO glass substrates (acetone, isopropanol, deionized water, and UV-ozone treatment) using a thermal evaporation system at pressure $< 2.0 \times 10^{-7}$ torr. The deposition rates of the organic and metal layers were controlled independently from 0.1 to 1 nm s$^{-1}$, while Liq was deposited at a rate 0.01 nm s$^{-1}$. The devices were encapsulated in a nitrogen-filled glove box prior to the measurements. The current density–voltage–luminance (*J*–*V*–*L*) characteristics and the EL spectra were measured using a programmable source meter (Keithley 2400, active area of devices 4 mm$^2$) and a spectrophotometer (Photo Research Spectrascan PR650). EQEs were estimated under the assumption of a Lambertian emission pattern. The lifetime measurements (LT$_{50}$) of devices were determined in constant current mode in a temperature controlled chamber (25°C). All organic films for optical and electrical characterization were thermally deposited onto the quartz or ITO substrates.

*Photophysical characterization*

The transient PL decays and PL spectra of the films and devices were analyzed using a $N_2$ laser (337 nm; Usho Optical Systems Co., Osaka, Japan) and a streak camera system (C10627; Hamamatsu Photonics, Shizuoka, Japan). The electrically pumped transient PL measurements were performed by combining and synchronizing quasi steady-state electrical pulses (pulse width 200 μs, repetition rate 20 Hz, DG645; Stanford Research Systems, Sunnyvale, CA) with the transient PL system excited at the middle of the voltage pulses.[23]

*Exciton quenching rate in EML*

The radiative lifetime of the dopant at constant current density of $J$=1.65 mA/cm$^2$ was measured by transient PL(t) of dopant emission in fresh and degraded devices using a streak camera system, and decreased from 1.46 $\mu s$ at t=0 to 1.16 $\mu s$ at $LT_{50}$ (238 hours), as shown in Fig. 4(a). The additional non-radiative decay rates originating from the energy transfer from the dopant exciton by the quencher, $k_{ET}^{DQ}[Q(t)]_{EML}$, (process ⑦ in Fig. 1. and Table 1) can be extracted from the exciton lifetimes and were linearly increased to $2 \times 10^5 s^{-1}$ at $LT_{50}$, as shown in Fig. 4(a). The PL intensities of dopant emission in the fresh and aged devices were measured, and also shown to decrease gradually to 57% of the fresh device at $LT_{50}$ (0.88 to 0.5). The reduction of PL intensity originates from two sources; one from the reduced energy transfer efficiency from host excitons to dopant (⑥ in Fig. 1. and Table 1) and the other from the ET of the dopant exciton to quencher (⑦ in Fig. 1. and Table 1) represented by

$$\frac{PL(t)}{PL(t=0)} = \frac{\eta_{ET}^{HD}(t)}{\eta_{ET}^{HD}(t=0)} \times \frac{q_{eff}(t)}{q_{eff}(t=0)}.$$ By combining the quantum efficiency of the dopant obtained by transient PL measurement (Fig. 4(a)) and the quantum efficiency of the EML measured by reduced PL intensity (Fig. 4(b)), we obtained the energy transfer rate from the host exciton to

quencher using equation ⑥ in Table 1. Fig. 4(b) (right scale) shows the energy transfer rate from the host excitons to quencher over time, which increased almost linearly to $4\times10^{10}s^{-1}$ at $LT_{50}$.

*Quencher analysis*

Desorption electrospray ionization mass spectroscopy (DESI-MS) was performed to confirm the degradation products of the electrically degraded blue PhOLED device. Laser desorption ionization (LDI) is a commonly used technique to analyze molecular fragmentation. However, there was a limitation to ionizing electron transporting materials in our system (Fig. S2.). The DESI-MS imaging method has been widely used in biotechnology,[31,32] but has not been applied to organic electronics. All MS experiments were performed using a mass spectrometer (Synapt G2-Si; Waters Corp., Milford, MA) with a 2nd generation 2D DESI ion source (Waters Corp.). The samples were placed on a 3D moving stage using double-sided tape and analyzed by DESI-MS in positive ion mode. Typical instrumental parameters used were 5 kV capillary voltage and 150°C source temperature. Acetonitrile:water (90:10) solution was used as spray solvent and delivered at a flow rate of 1 µL/min. Leucine enkephalin (0.2 ng/µL) was added to the solution as lock-spray solution. Mass spectra were acquired as full scans in positive mode over the mass range from m/z=200 to 1700. The sprayer to surface distance was 1.0–1.5 mm, the sprayer to inlet distance was 3–5 mm, and the incident spray was set at 60°. To acquire DESI-MS images, the samples were scanned in horizontal rows separated by 100 to 100 µm vertical steps until the specified area of the sample was analyzed. The lines were scanned at a constant velocity of 100 m/s and the scan time was set to 0.985 s. A spatial resolution (pixel size) in the range of 100 to 100 µm could be achieved under these conditions. Scan area was defined by 10 mm in length and 3 mm in width, covering fresh and aged pixels simultaneously for quantitative and qualitative analyses. Data were acquired and processed using Masslynx

4.0 software and HDI 1.4 software (Waters Corp.). Multivariate analysis was performed using Progenesis QI 2.4 software (Waters Corp.) defining four regions of interest in each active area. An ultra-high resolution mass spectrometer (MALDI solariX FT-ICR 9.4T; Bruker, Karlsruhe, Germany) was used for the analysis. The mass spectrometer was operated in positive ion mode with m/z=1200 m/z with a resolution of 400000 at m/z 200. Two thousand laser shots (2 kHz via Bruker proprietary Smartbeam II MALDI source) were automatically acquired for each spectrum.

## ACKNOWLEDGEMENTS

We thank Sangmi Oh from Waters Korea for desorption electrospray ionization mass spectroscopy (DESI-MS) measurements. This work was supported by National Research Foundation. Grant Numbers: 2017R1A2A1A05022985, 2018R1A2A1A05078734 funded by the Ministry of Science and ICT (MSIT) and the Samsung Advanced Institute of Technology, Samsung Electronics Co., Ltd.

## APPENDIX A: APPLICATION OF VOLTAGE RISE MODEL

Consider the device structure shown below where $x$ is defined from HTL to ETL.

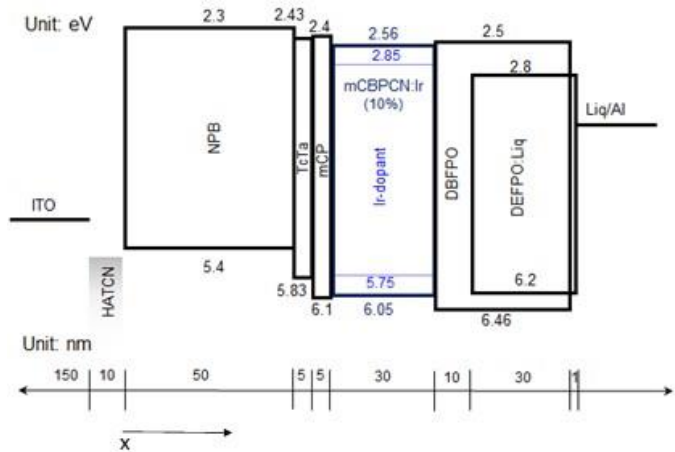

The generation rate of quencher density can be represented as follows:

$$\frac{d[Q(x,t)]}{dt} = k_{QF}^{P}[P(x,t)] + \left(k_{QF}^{E} + k_{QF}^{EP}[P(x,t)]\right)[N(x,t)] + k_{QF}^{EE}[N(x,t)]^2 + k_{QF}^{imp}[P(x,t)][A(x,t)]$$

Therefore, the density of quencher under constant current operation is represented as:

$$[Q(x,t)] = k_{QF}^{P}[P(x)]t + \left(k_{QF}^{E} + k_{QF}^{EP}[P(x)]\right)\int_{0}^{t}[N(x,t)]dt \\ + k_{QF}^{EE}\int_{0}^{t}[N(x,t)]^{2}dt + [A(x)]_{0}\left(1-\exp(-k_{QF}^{imp}[p]t)\right)$$

The quencher density at position $x$, time $t$ can be expressed by $[Q(x,t)] = [Q(t)] \cdot g(x)$ if the polaron, exciton, and impurity distributions do not change over time. The normalized quencher distribution function, $g(x)$, satisfies the condition, $\int_{0}^{x}g(x)dx = 1$. If the total exciton density in the device during electrical operation is assumed to be proportional to that in the emitting layer $[\tilde{N}(t)]$, the voltage rise over time is represented as follows:

$$\Delta V(t) = \frac{e}{\varepsilon\varepsilon_{0}}\int_{0}^{x} f \cdot x \cdot [Q(x,t)]dx$$

$$\Delta V(t) = \frac{e}{\varepsilon\varepsilon_{0}} f \begin{pmatrix} k_{QF}^{P}[P]t\int_{0}^{x}x \cdot g_{P}(x)dx \\ + k_{QF}^{E}\int_{0}^{t}[N(t)]dt \cdot \int_{0}^{x}x \cdot g_{N}(x)dx \\ + k_{QF}^{EP}[P]\int_{0}^{t}[N(t)]dt \cdot \int_{0}^{x}x \cdot g_{NP}(x)dx \\ + k_{QF}^{EE}\int_{0}^{t}[N(t)]^{2}dt \cdot \int_{0}^{x}x \cdot g_{NN}(x)dx \\ + [A]_{0}\left(1-\exp(-k_{En}t)\right) \cdot \int_{0}^{x}x \cdot g_{En}(x)dx \end{pmatrix}.$$

Here, $g_{P}(x)$ is the polaron distribution in the device, which can be calculated by drift-diffusion simulation. $g_{N}(x)$ is the exciton distribution in the device. The exciton profile in EML is obtained by the sensing layer method. $g_{NP}(x)$ is the overlap of the polaron profile and exciton profile in the device. $g_{NN}(x)$ is expressed as the square of the exciton profile in the device. $g_{En}(x)$ is the profile of initial impurity density in the device. Then, $\Delta V(t)$ under constant current can be expressed as:

$$\Delta V(t) = \left[ C_1 t + C_2 \int_0^t [N(t)] dt + C_3 \int_0^t [N(t)]^2 dt + A^*\left(1 - \exp(-k_{QF}^{imp}[p]t)\right) \right],$$

where

$$C_1 = k_{QF}^P [P] \int_0^x x \cdot g_P(x) dx,$$

$$C_2 = k_{QF}^E \cdot \int_0^x x \cdot g_N(x) dx + k_{QF}^{EP}[P] \int_0^x x \cdot g_{NP}(x) dx,$$

$$C_3 = k_{QF}^{EE} \cdot \int_0^x x \cdot g_{NN}(x) dx, \text{ and}$$

$$A^* = [A]_0 \cdot \int_0^x x \cdot g_{En}(x) dx.$$

Thus, the voltage rise under electrical operation can be fitted with four fitting parameters representing the polaron-induced mechanism ($C_1$), exciton-related first-order reaction ($C_2$), exciton-related second-order reaction, ($C_3$) and impurities ($C_4$).

**APPENDIX B: EXCITON DENSITY, [N(t)]**

The initial density of the exciton, $[N]_0$ is represented as

$$[D]_0 = \eta_{EF,\text{int}} \cdot (k_{rec}^H \eta_{ET,\text{int}}^{HD} + k_{rec}^D) \cdot \tau_D$$

The changes of the exciton density over time is

$$\frac{d[N(t,t')]}{dt'} = \eta_{EF}(t)\left(k_{rec}^H \cdot \eta_{ET}^{HD}(t) + k_{rec}^D\right) - (k_r^D + k_{nr}^D(t))[N(t,t')]$$

where, $t$ is time with unit of hours and $t'$ is time with unit of $\sim \mu s$.

At steady-state for short time range ($\sim \mu s$),

$$[N(t)] = \frac{\eta_{EF}(t)\left(k_{rec}^H \cdot \eta_{ET}^{HD}(t) + k_{rec}^D\right)}{k_r^D + k_{nr}^D(t)} \quad \text{(unit: } \frac{\#}{cm^3} = \frac{cm^{-3}s^{-1}}{s^{-1}}\text{)}$$

[1] S.-Y. Kim, W.-I. Jeong, C. Mayr, Y.-S. Park, K.-H. Kim, J.-H. Lee, C.-K. Moon, W. Brütting, and J.-J. Kim, Organic light-emitting diodes with 30% external quantum efficiency based on a horizontally oriented emitter, Adv. Funct. Mater. **23**, 3896 (2013).

[2] K.-H. Kim, J.-L. Liao, S. W. Lee, B. Sim, C.-K. Moon, G.-H. Lee, H. J. Kim, Y. Chi, and J.-J. Kim, Crystal organic light-emitting diodes with perfectly oriented non-doped Pt-based emitting layer, Adv. Mater. **28**, 2526 (2016).

[3] D. H. Ahn, S. W. Kim, H. Lee, I. J. Ko, D. Karthik, J. Y. Lee, and J. H. Kwon, Highly efficient blue thermally activated delayed fluorescence emitters based on symmetrical and rigid oxygen-bridged boron acceptors, Nature Photonics, https://doi.org/10.1038/s41566-019-0415-5 (2019).

[4] H. Shin, Y. H. Ha, H.-G. Kim, R. Kim, S.-K. Kwon, Y.-H. Kim, and J.-J. Kim, Controlling horizontal dipole orientation and emission spectrum of Ir complexes by chemical design of ancillary ligands for efficient deep-blue organic light-emitting diodes. Adv. Mater. **31**, 1808102 (2019).

[5] S. Scholz, D. Kondakov, B. Lussem, and K. Leo, Degradation mechanisms and reactions in organic light-emitting devices. Chem. Rev. **115**, 8449 (2015).

[6] N. C. Giebink, B. W. D'Andrade, M. S. Weaver, P. B. Mackenzie, J. J. Brown, M. E. Thompson, and S. R. Forrest, Intrinsic luminance loss in phosphorescent small-molecule organic light emitting devices due to bimolecular annihilation reactions, J. Appl. Phys. **103**, 044509 (2008).

[7] H. Yamamoto, J. Brooks, M. S. Weaver, J. J. Brown, T. Murakami, and H. Murata, Improved initial drop in operational lifetime of blue phosphorescent organic light emitting device fabricated under ultra high vacuum condition, Appl. Phys. Lett. **99**, 033301 (2011).

[8] J. Lee, C. Jeong, T. Batagoda, C. Coburn, M. E. Thompson, and S. R. Forrest, Hot excited state management for long-lived blue phosphorescent organic light-emitting diodes, Nat. Commun. **8**, 15566 (2017).

[9] S. Kim, H. J. Bae, S. Park, W. Kim, J. Kim, J. S. Kim, Y. Jung, S. Sul, S.-G. Ihn, C. Noh, S. Kim, and Y. You, Degradation of blue-phosphorescent organic light-emitting devices involves exciton-induced generation of polaron pair within emitting layers, Nat. Commun. **9**, 1211 (2018).

[10] P. Heimel, A. Mondal, F. May, W. Kowalsky, C. Lennartz, D. Andrienko, and R. Lovrincic, Unicolored phosphor-sensitized fluorescence for efficient and stable blue OLEDs, Nat. Commun. **9**, 4990 (2018).

[11] N. C. Giebink, B. W. D'Andrade, M. S. Weaver, J. J. Brown, and S. R. Forrest, Direct evidence for degradation of polaron excited states in organic light emitting diodes, J. Appl. Phys. **105**, 124514 (2009).

[12] Q. Wang, B. Sun, and H. Aziz, Exciton-polaron-induced aggregation of wide-bandgap materials and its implication on the electroluminescence stability of phosphorescent organic light-emitting devices, Adv. Funct. Mater. **24**, 2975 (2014).

[13] D. Y. Kondakov, W. C. Lenhart, and W. F. Nichols, Operational degradation of organic light-emitting diodes: Mechanism and identification of chemical products, J. Appl. Phys. **101**, 024512 (2007).

[14] A. S. D. Sandanayaka, T. Matsushima, and C. Adachi, Degradation mechanisms of organic light-emitting diodes based on thermally activated delayed fluorescence molecules, J. Phys. Chem. C **119**, 23845 (2015).

[15] H. Yu, Y. Zhang, Y. J. Cho, and H. Aziz, Exciton-induced degradation of carbazole-based host materials and its role in the electroluminescence spectral changes in phosphorescent organic light emitting devices with electrical aging, ACS Appl. Mater. Interfaces **9**, 14145 (2017).

[16] T. D. Schmidt, L. Jager, Y. Noguchi, H. Ishii, and W. Brutting, Analyzing degradation effects of organic light-emitting diodes via transient optical and electrical measurements, J. Appl. Phys. **117**, 215502 (2015).

[17] K. W. Hershey, J. S. Bangsund, G. Qian, and R. J. Holmes, Decoupling degradation in exciton formation and recombination during lifetime testing of organic light-emitting devices, Appl. Phys. Lett. **111**, 113301 (2017).

[18] J. S. Bangsund, K. W. Hershey, and R. J. Holmes, Isolating degradation mechanisms in mixed emissive layer organic light-emitting devices, ACS Appl. Mater. Interfaces **10**, 5693 (2018).

[19] Q. Niu, C.-J. AQ. Niu, C.-J. A.H. Wetzlaer, P. W. M. Blom, and N. I. Craciun, Modeling of Electrical Characteristics of Degraded Polymer Light-Emitting Diodes, Adv. Electron. Mater. **26**, 1600103 (2016).

[20] M. A. Baldo, C. Adachi, and S. R. Forrest, Transient analysis of organic electrophosphorescence. II. Transient analysis of triplet-triplet annihilation, Phys. Rev. B **62**, 16 (2000).

[21] S. Reineke, K. Walzer, and K. Leo, Triplet-exciton quenching in organic

phosphorescent light-emitting diodes with Ir-based emitters, Phys. Rev. B **75**, 125328 (2007).

[22] N. C. Giebink, and S. R. Forrest, Quantum efficiency roll-off at high brightness in fluorescent and phosphorescent organic light emitting diodes, Phys. Rev. B, **77**, 235215 (2008).

[23] B. Sim, C.-K. Moon, K.-H. Kim, and J.-J. Kim, Quantitative analysis of the efficiency of OLEDs, ACS Appl. Mater. Interfaces **8**, 33010 (2016).

[24] S. Nowy, B. C. Krummacher, J. Frischeisen, N. A. Reinke, and W. Brutting, Light extraction and optical loss mechanisms in organic light-emitting diodes: influence of the emitter quantum efficiency, J. Appl. Phys. **104**, 123109 (2008).

[25] M. Murgia, R. H. Michel, G. Ruani, W. Gebauer, O. Kapousta, R. Zamboni, and C. Taliani, In-situ characterization of the oxygen induced changes in a UHV grown organic light-emitting diode, Synth. Met. **102**, 1095 (1999).

[26] T. Ikeda, H. Murata, Y. Kinoshita, J. Shike, Y. Ikeda, and M. Kitano, Enhanced stability of organic light-emitting devices fabricated under ultra-high vacuum condition, Chem. Phys. Lett. **426**, 111 (2006).

[27] S.-G. Ihn, N. Lee, S. O. Jeon, M. Sim, H. Kang, Y. Jung, D. H. Huh, Y. M. Son, S. Y Lee, M. Numata, H. Miyazaki, R. G. Bombarelli, J. A. Iparraguirre, T. Hizel, A. A. Guzik, S. Kim, and S. Lee, An alternative host material for long-lifespan blue organic light-emitting diodes using thermally activated delayed fluorescence, Adv. Sci. 1600502 (2017).

[28] Y. Sato, S. Ichinosawa, and H. Kanai, Operation characteristics and degradation of organic electroluminescent devices, IEEE J. Sel. Top. Quantum Electron. **4**, 40 (1998).

[29] R. C. Kwong, M. R. Nugent, L. Michalski, T. Ngo, K. Rajan, Y.-J. Tung, M. S. Weaver, T. X. Zhou, M. Hack, M. E. Thompson, S. R. Forrest, and J. J. Brown, High operational stability of electrophosphorescent devices. Appl. Phys. Lett. **81**, 162 (2002).

[30] C. Féry, B. Racine, D. Vaufrey, H. Doyeux, and S. Cina, Physical mechanism responsible for the stretched exponential decay behavior of aging organic light-emitting diodes. Appl. Phys. Lett. **87**, 213502 (2005).

[31] Y. Dong, B. Li, S. Malitsky, I. Rogachev, A. Aharoni, F. Kaftan, A. Svatos, and P. Franceschi, Sample preparation for mass spectrometry imaging of plant tissues:a review, Front. Plant Sci. **7**, 60 (2016).

[32] J. Xu, Y. Chu, B. Liao, S. Xiao, Q. Yin, R. Bai, H. Su, L. Dong, X. Li, J. Qian, J. Zhang, Y. Zhang, X. Zhang, M. Wu, J. Zhang, G. Li, L. Zhang, Z. Chang, Y. Zhang, Z. Jia, Z. Liu, D. Afreh, R. Nahurira, L. Zhang, R. Cheng, Y. Zhu, G. Zhu, W. Rao, C. Zhou, L. Qiao, Z. Huang, Y.-C. Cheng, and S. Chen, Panax ginseng genome examination for ginsenoside biosynthesis, Giga Science. **6**, 11 (2017).

[33] O. Molt, C. Lennartz, G. Wagenblast, E. Fuchs, N. Langer, C. Schidknecht, K. Dormann, S. Watanabe, T. Schaefer, H. Wolleb, T. Marina, F. Duarte, S. Metz, and P. Murer (UDC Ireland Ltd.), Metal complexes comprising dazabenzmidazolocarbene ligands and the use thereof in OLEDs, US20130032766A1 (2013).

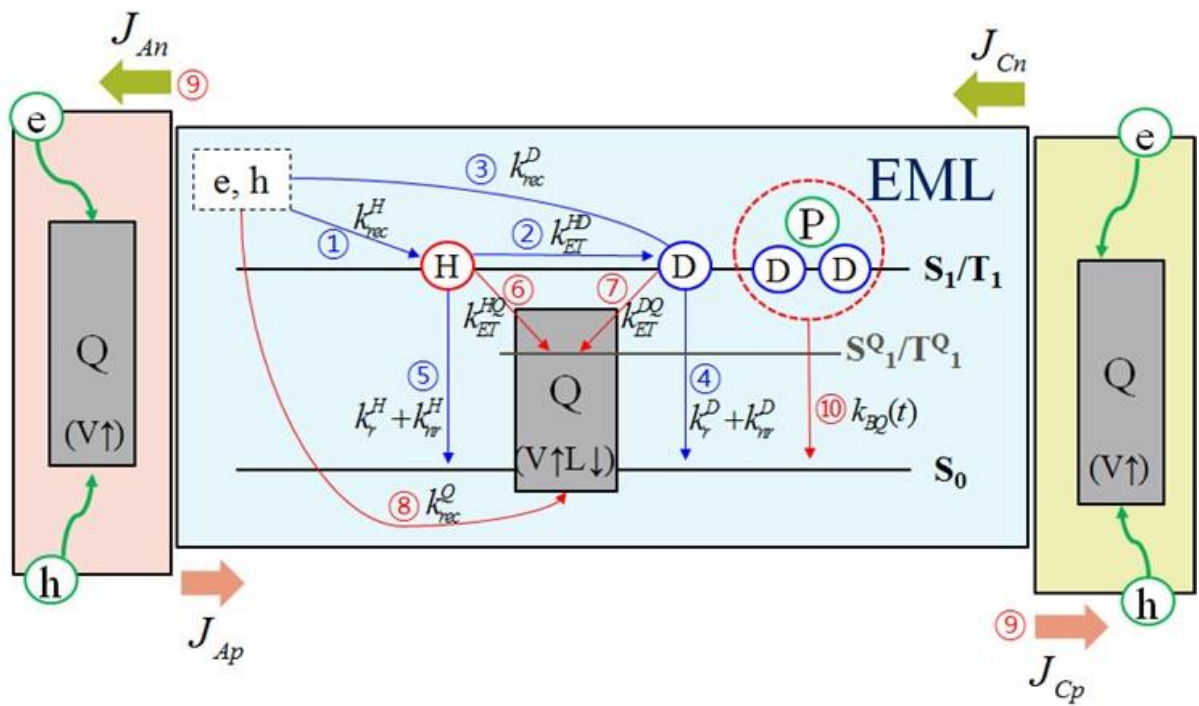

**FIG. 1.** Schematic diagram of the exciton generation and annihilation processes. ① Charge recombination at host. ② Energy transfer from host exciton to dopant. ③ Charge recombination at dopant. ④ Radiative and non-radiative decay of dopant exciton. ⑤ Radiative and non-radiative decay of host exciton. ⑥ Energy transfer from host exciton to quencher. ⑦ Energy transfer from dopant exciton to quencher. ⑧ Charge recombination at quencher. ⑨ Charge leakage from EML. ⑩ Biparticle interaction of dopant exciton.

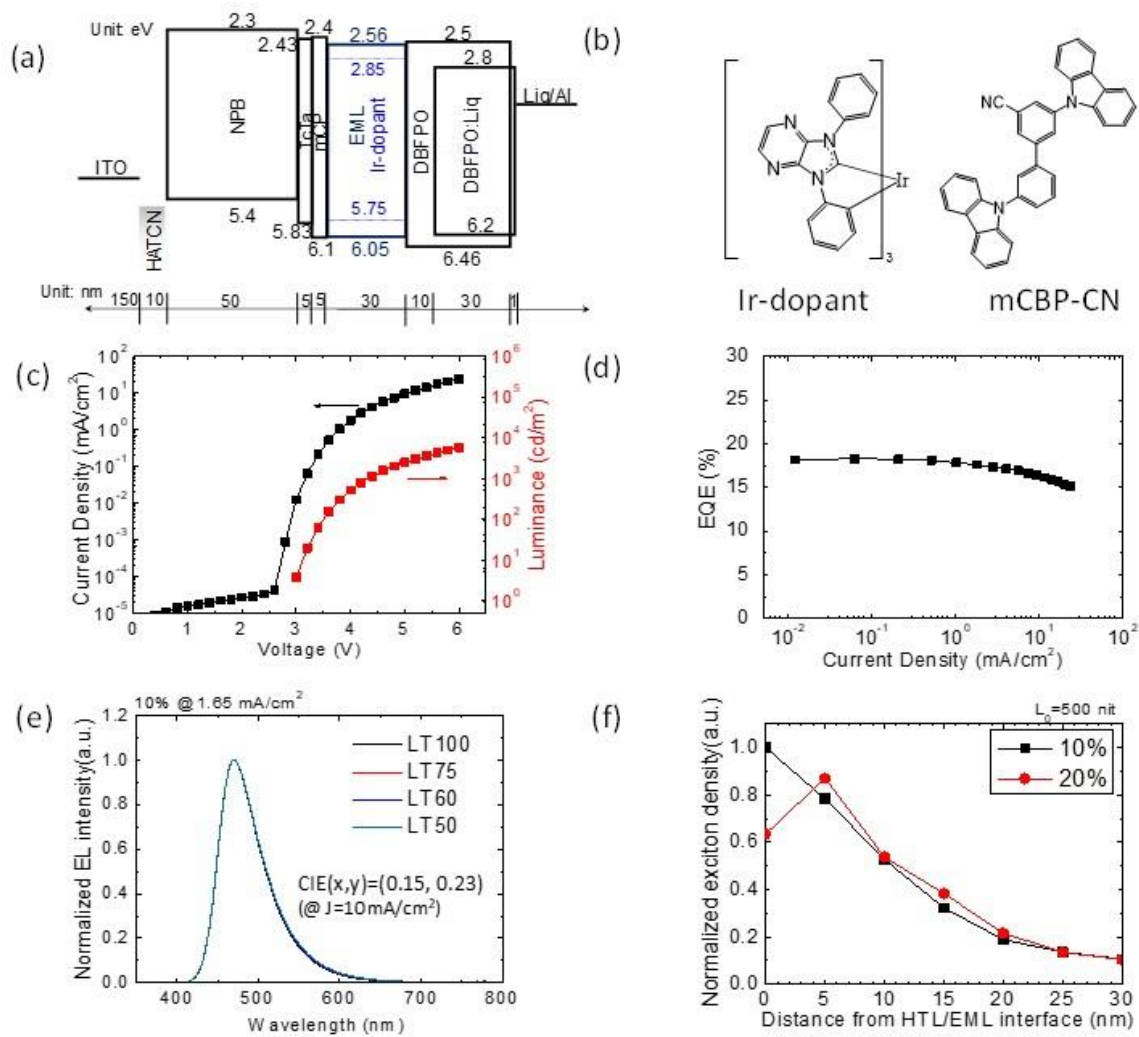

**FIG. 2.** Electrical characteristics of the blue phosphorescent device used in our model. (a) Schematic device structure with energy (eV) diagram. (b) Chemical structures of dopant and host. (c) Current density–voltage–luminance (*J–V–L*) characteristics. (d) EQE as a function of current density (*EQE–J*). (e) Normalized EL spectra over time. The peak wavelength was constant during degradation. (f) Exciton profiles at 500 cd/m$^2$ determined by the sensing layer method.

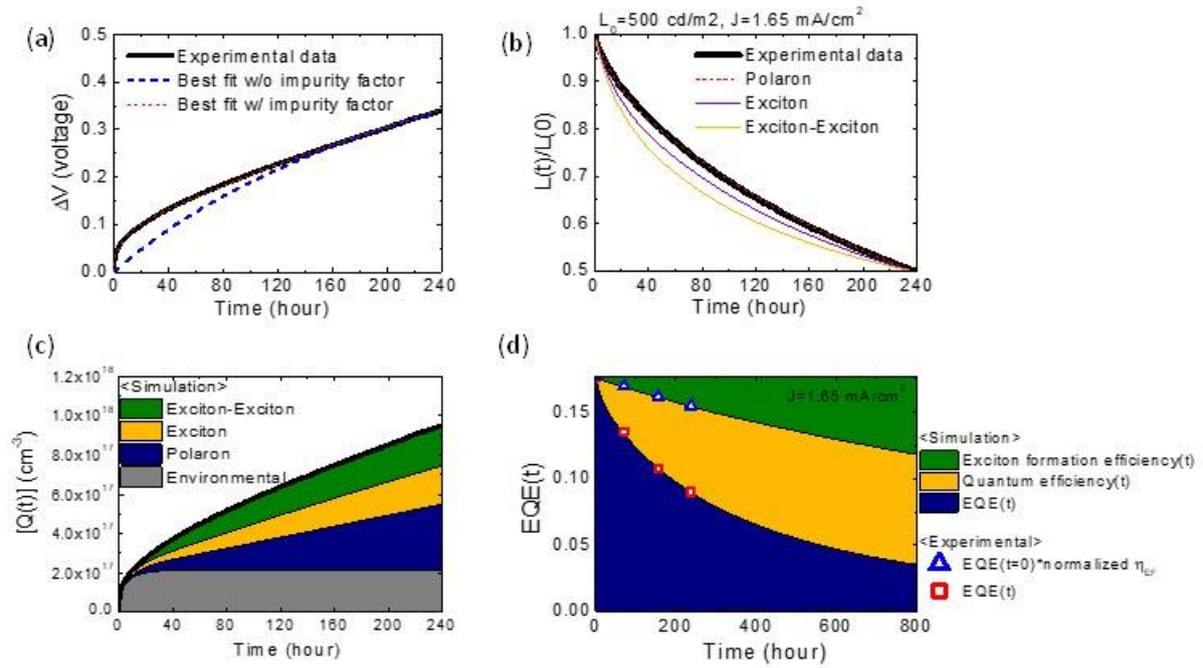

**FIG. 3.** Analysis of the electrical characteristics of the blue phosphorescent device used in our model over time. (a) The experimental driving voltage change (black line) was fitted by a voltage rise model, with (red dotted line) and without impurity factor (blue dashed line). (b) Luminance and EQE change as a function of time at initial luminance of 500 cd/m$^2$ compared with the results of fitting based on different mechanisms where the impurity effect was included (c) Quencher densities generated over time by different mechanisms are indicated by differently colored areas. (d) The contributions of the reduced exciton formation efficiency (green) and the reduced EQE (yellow area) to EQE loss were compared with experimental data (dots) extracted from LT$_{100}$ (fresh), LT$_{75}$, LT$_{60}$, and LT$_{50}$ devices..

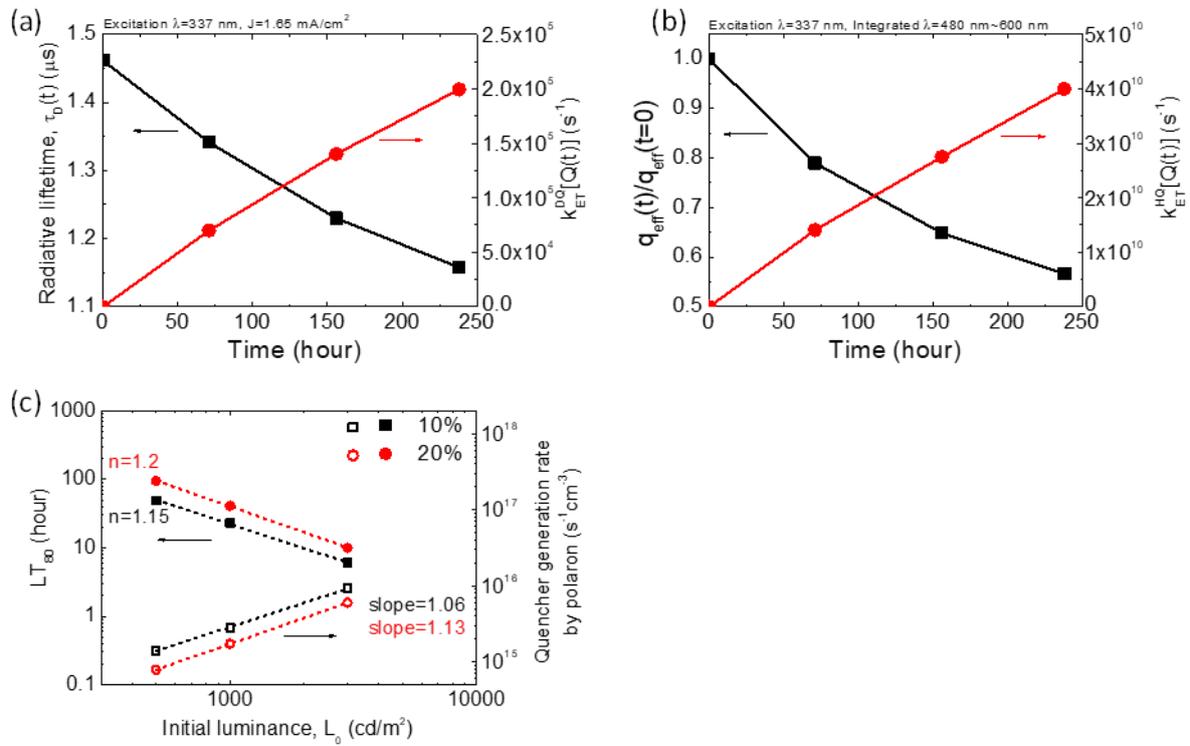

**FIG. 4.** Exciton quenching rates due to the quenchers. (a) Quenching rates of dopant excitons through energy transfer from the dopant to quencher (red circles) extracted from the measured exciton lifetime (black square) using transient PL data. (b) Quenching rate of host exciton through energy transfer from host exciton to quencher (red circles) obtained from the reduced $q_{eff}(t)$ (black squares) and the integrated PL intensity after degradation of the device. (c) The operation time for $LT_{80}$ (left scale, closed symbols) and the quencher generation rates in the EML (right scale, open symbols) under various initial luminance conditions.

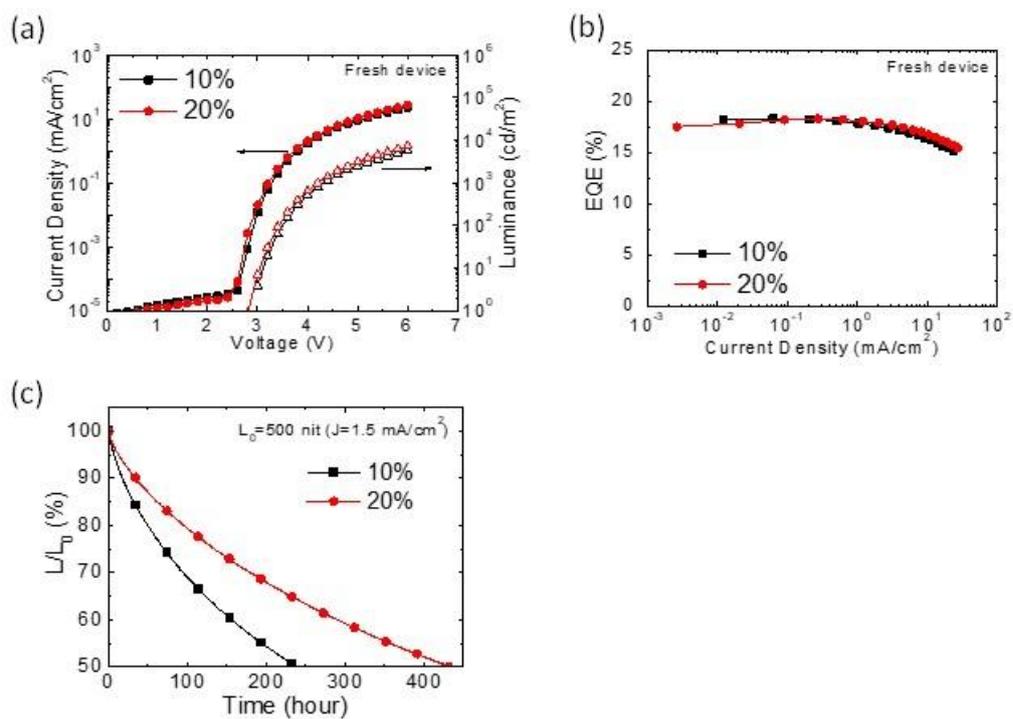

**FIG. 5.** Electrical characteristic of 20% doped device (red). The data of 10% doped device are included to be compared (black). (a) (*J–V–L*) characteristics, (b) EQE as a function of current density (*EQE–J*). (c) Luminance change as a function of time at initial luminance of 500 cd/m$^2$. The lifetime of the 20% doped device is two time longer with LT$_{50}$ of 431 hours at an initial luminance of 500 cd/m$^2$.

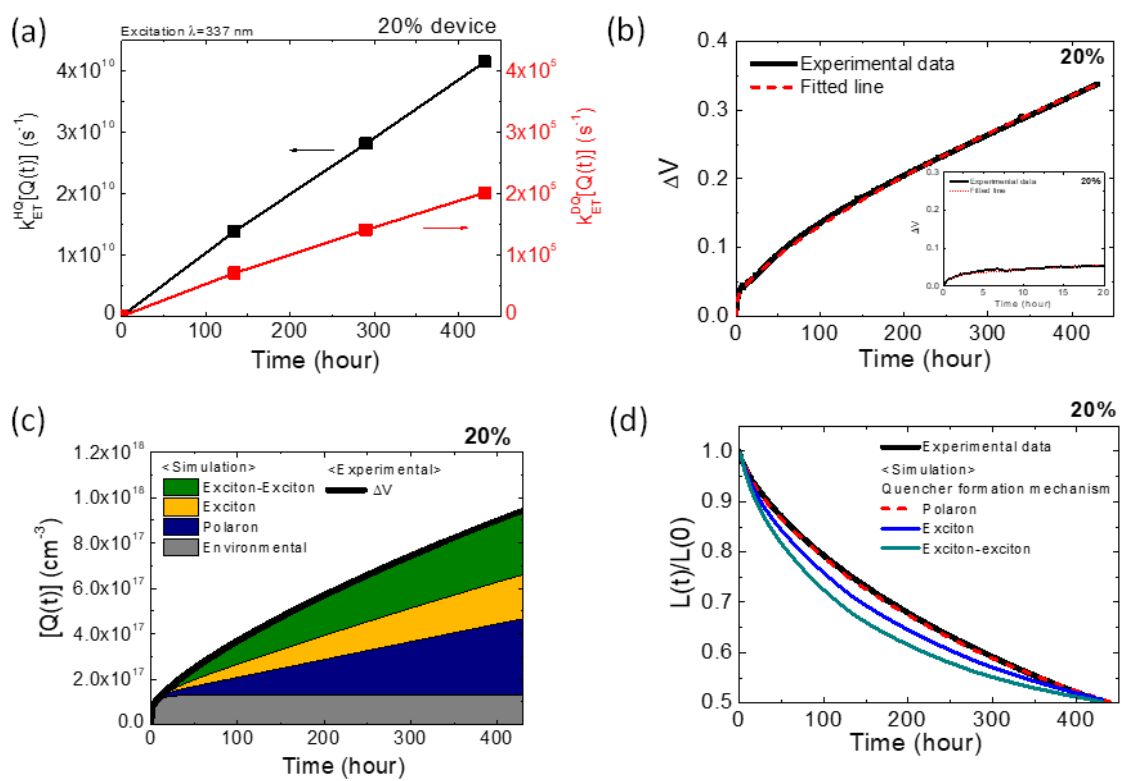

**FIG. 6.** Analysis results of 20% doped device. (a) Exciton quenching rate of 20% doped device. (b) Fitting results of voltage rise model. (c) Quencher generation rate and density. (d) Prediction by luminance loss model for exciton quenching rate of 20% doped device.

.

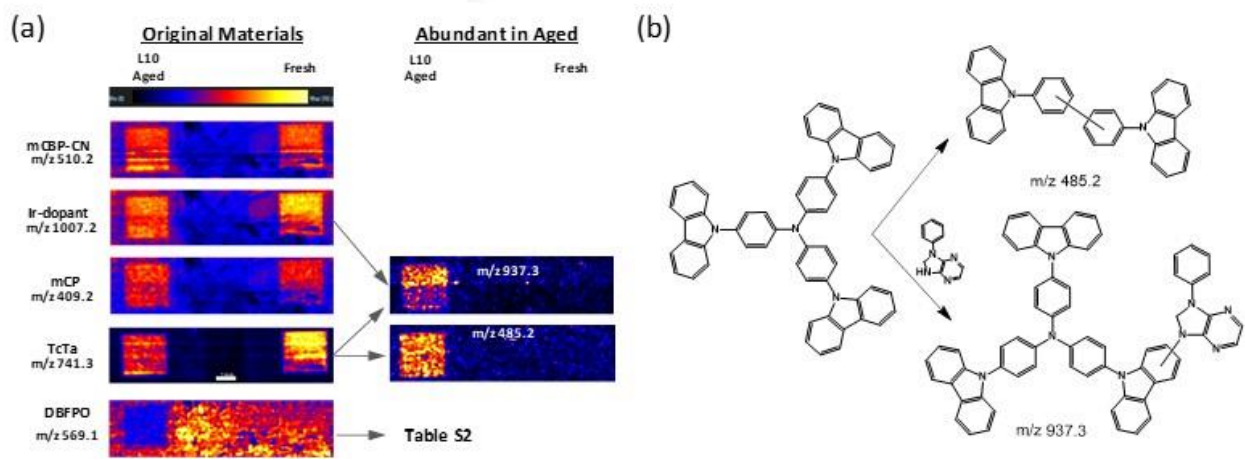

**FIG. 7.** Desorption electrospray ionization mass spectrometry (DESI-MS) of electrically degraded device. (a) DESI-MS images of host, dopant, mCP, TCTA, DBFPO (left column), and degradation products (right column), and (b) tentative molecular structures and possible reaction pathways to form m/z=485.2 and 937.3. The intensities of the colors represent the relative amounts of materials.

**Table I.** Equations, units, measurable parameters, and experiments to determine the parameters used in the model. The numbers in the first column correspond to the processes in FIG. 1.

| # | Equation | Unit | Measurable parameter | Experimental |
|---|---|---|---|---|
| ① | $k_{rec}^{H} = r_L = \dfrac{e \cdot (\mu_n + \mu_p)}{\varepsilon \cdot \varepsilon_0} \cdot n \cdot p$ | $cm^{-3}s^{-1}$ | $\mu_n, \mu_p$ *(n, p) | Mobility measurement of EML using single carrier devices *(calculated by drift-diffusion simulation) |
| ② | $k_{ET}^{HD} = \dfrac{1}{\tau_{EML}^0} - \dfrac{1}{\tau_H^0} = (k_r^H + k_{nr}^H + k_{ET}^{HD}) - (k_r^H + k_{nr}^H)$ | $s^{-1}$ | $\tau_{EML}^0, \tau_H^0$ | Transient PL of host emission in thin films (host and host:dopant films at λ=host emission range) |
| ③ | $k_{rec(h)}^{D} = r_T = \dfrac{e \cdot \mu_n}{\varepsilon \cdot \varepsilon_0} \cdot n \cdot p_t$ (hole trap) <br><br> $k_{rec(e)}^{D} = r_T = \dfrac{e \cdot \mu_p}{\varepsilon \cdot \varepsilon_0} \cdot p \cdot n_t$ (electron trap) | $cm^{-3}s^{-1}$ | $\mu_n, \mu_p$ *(n, p, n$_t$, p$_t$) | from ① *(calculated by drift-diffusion simulation) |
| ④ | $\tau_D = \dfrac{1}{Fk_r^D + k_{nr,int}^D}$ | s | $\tau_D$ | Transient PL of dopant emission in fresh device |
| ⑤ | $\tau_H^0 = \dfrac{1}{k_r^H + k_{nr,int}^H}$ | s | $\tau_H^0$ | Transient PL of host emission in film |
| ⑥ | $\eta_{ET}^{HD}(t) = \dfrac{k_{ET}^{HD}}{k_{ET}^{HD} + Fk_r^H + k_{nr,int}^H + k_{ET}^{HQ}[Q(t)]_{EML}}$ | - | $\eta_{ET}^{HD}(t), \tau_H$ | Transient PL(t) and PL(t) intensity of dopant emission in degraded device, and ②, ⑤ |
| ⑦ | $k_{ET}^{DQ}[Q(t)]_{EML} = \dfrac{1}{\tau_D(t)} - \dfrac{1}{\tau_D}$ | $s^{-1}$ | $\tau_D(t), \tau_D$ | Transient PL of dopant emission in degraded device, and ④ |
| ⑧ | $k_{rec(h)}^{Q} = \dfrac{e}{\varepsilon \varepsilon_0} \mu_n \cdot [Q(t)]_{EML} \cdot (1-f) \cdot n$ (hole trap) <br><br> $k_{rec(e)}^{Q} = \dfrac{e}{\varepsilon \varepsilon_0} \mu_p \cdot [Q(t)]_{EML} \cdot f \cdot p$ (electron trap) | $cm^{-3}s^{-1}$ | $\mu_n, \mu_p$, $[Q(t)]_{EML}$ | Time dependent driving voltage ($\Delta V(t)$), and ③ |
| ⑨ | $\eta_{EF}(J,t) = \left(\dfrac{J_{Cn}(t) - J_{An}(t)}{J}\right) \cdot \left(\dfrac{k_{rec}^H + k_{rec}^D}{k_{rec}^H + k_{rec}^D + k_{rec}^Q(t)}\right)$ | - | $k_{rec}^H, k_{rec}^D, k_{rec}^Q(t)$ | by ①, ③, ⑧ |
| ⑩ | $k_{BQ}(J,t) = \dfrac{1}{\tau_D(J,t)} - \dfrac{1}{\tau_D(J=0,t)}$ | $s^{-1}$ | $\tau_D(J,t)$ $\tau_D(J=0,t)$ | Electrically pumped transient PL of dopant emission in fresh and degraded device at current density of J |

**Table II.** Quencher generation rate constants from fit by voltage rise model.

| | $k_{QF}^{P}[p]$ $(cm^{-3}s^{-1})$ | $k_{QF}^{E}+k_{QF}^{Ep}[p]$ $(s^{-1})$ | $k_{QF}^{EE}$ $(cm^{3}s^{-1})$ | $k_{QF}^{imp}[p]$ $(s^{-1})$, | $[A]_0$ $(cm^{-3})$ |
|---|---|---|---|---|---|
| 10% | $1.4\times10^{15}$ | $2.5\times10^{-1}$ | $8.0\times10^{-17}$ | $1.6\times10^{-1}$ | $2.1\times10^{17}$ |
| 20% | $7.84\times10^{14}$ | $1.24\times10^{-1}$ | $5.87\times10^{-17}$ | $9.1\times10^{-2}$ | $1.3\times10^{17}$ |

**Table III.** Experimentally measured parameters used for device simulation

| Parameter | Symbol | Value |
|---|---|---|
| Effective radiative decay rate of dopant | $Fk_r^D$ | $6.8 \times 10^5$ s |
| Non-radiative decay rate of dopant | $k_{nr,int}^D$ | $9.2 \times 10^4$ s |
| Radiative decay rate of host | $k_r^H$ | $1.6 \times 10^7$ s |
| Non-radiative decay rate of host | $k_{nr,int}^H$ | $1 \times 10^8$ s |
| Host to dopant energy transfer rate | $k_{ET}^{HD}$ | $6 \times 10^{10}$ s |
| Recombination rate at host | $k_{rec}^H$ | $4 \times 10^{20} cm^{-3} s^{-1}$ |
| Recombination rate at dopant | $k_{rec}^D$ | $1.5 \times 10^{20} cm^{-3} s^{-1}$ |
| Ratio of exciton formation at host[a] (optical excitation at 337 nm) | $\alpha_{PL}$ | 0.91 |
| Ratio of exciton formation at host[b] (electrical excitation) | $\alpha_{EL}$ | 0.7 |
| Electron mobility in EML | $\mu_n$ | $2 \times 10^{-9} cm^2/V \cdot s$ |
| Hole mobility in EML | $\mu_h$ | $5 \times 10^{-8} cm^2/V \cdot s$ |
| Electron density[c] | $n$ | $3 \times 10^{17} cm^{-3}$ |
| Hole density[c] | $p$ | $5 \times 10^{16} cm^{-3}$ |
| Trapped hole density[c] | $p_t$ | $5 \times 10^{17} cm^{-3}$ |
| Width of emission zone[d] | $L$ | $15 nm$ |
| Relative permittivity | $\varepsilon$ | 3.5 |

a: Calculated by ratio of absorption coefficient

b: Calculated by ratio of recombination rate of host and dopant in the case of hole trap

c: Assumed based on the drift diffusion simulation

d: Measured by the sensing layer method

# Supporting Information

**A Comprehensive Model of the Degradation of Organic Light-Emitting Diodes and Application for Efficient Stable Blue Phosphorescent Devices with Reduced Influence of Polarons**

Bomi Sim[1], Jong Soo Kim[3], Hyejin Bae[3], Sungho Nam[3], Eunsuk Kwon[3], Ji Whan Kim[3], Hwa-Young Cho[3], Sunghan Kim[3*], Jang-Joo Kim[1,2*]

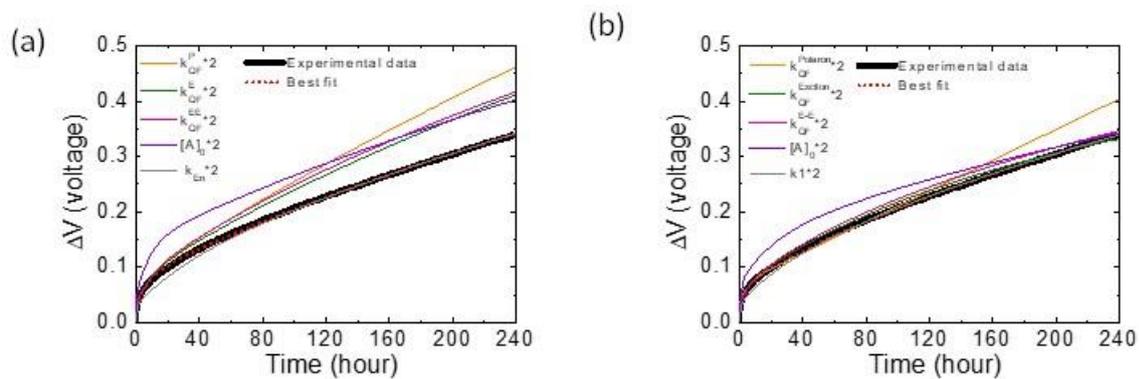

**FIG. S1.** (a) Each parameter is doubled compared to the best fit in FIG. 3(a), while other parameters are fixed. (b) Each parameter is doubled compared to the best fit in FIG 3(a), and other parameters are changed to be fitted in the experimental data.

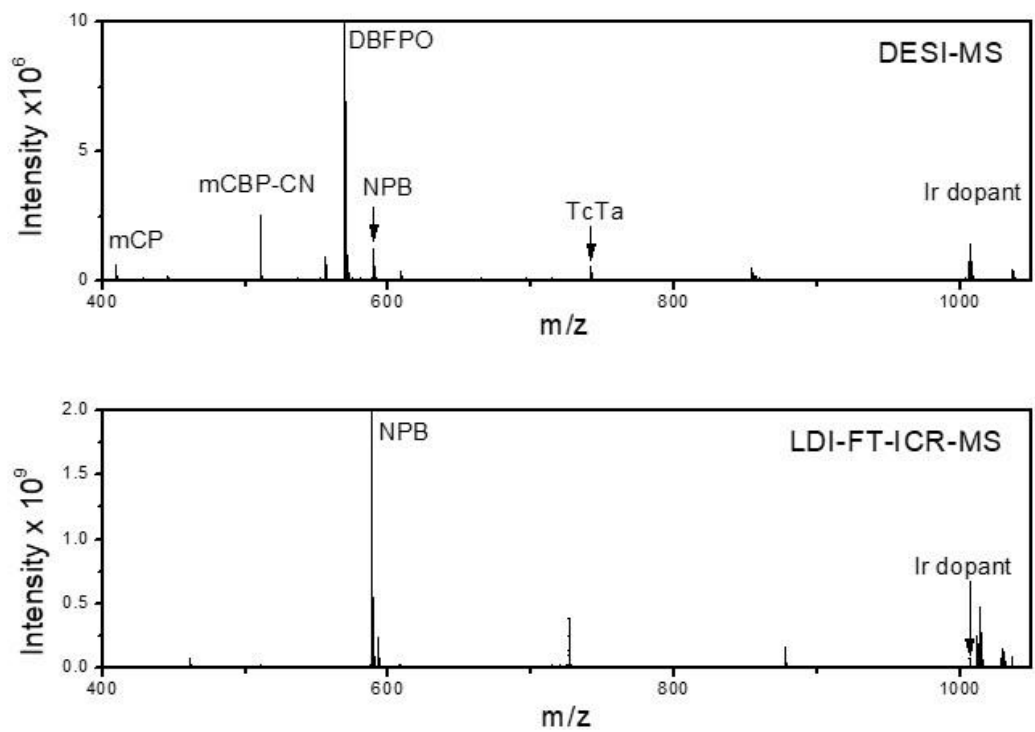

**FIG. S2.** Comparison of (a) LDI-MS and (b) DESI-MS.

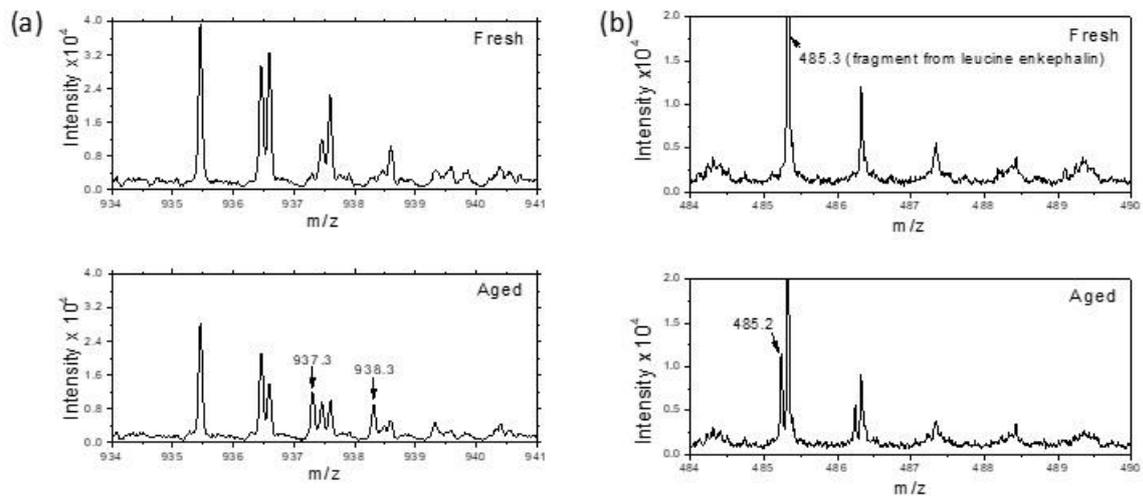

**FIG. S3.** Mass spectra of degradation products compared to those of fresh pixels (m/z 937.3 and 485.2).

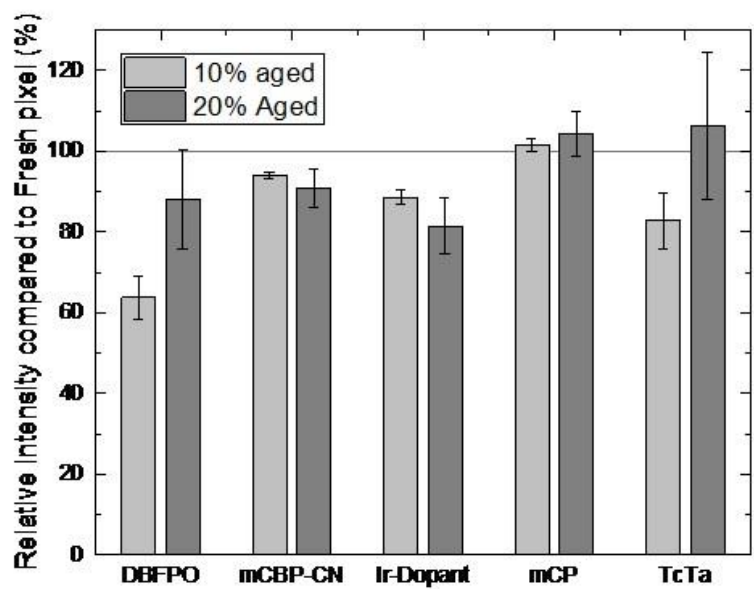

**FIG. S4.** Relative intensity compared to fresh pixels of compounds comprising OLEDs in this study. The 100% reference line is shown in the figure.

**Table SI.** DESI-MS images of distinguishable compounds in LT$_{10}$ and fresh pixel, and tentative molecular structure of those mass

| Material | m/z A/F fold* | DESI-MS images (L10 Aged / Fresh) | Tentative structure |
|---|---|---|---|
| NPB | 589.3 | 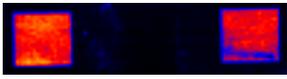 | 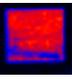 |
| DBFPO | 569.1 | 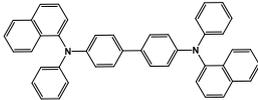 | 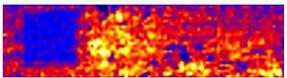 |
| degradation product | 369.1 9.3 | 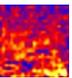 | 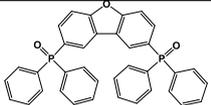 |
| | 493.1 inf. | 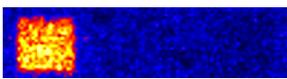 | 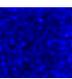 − Phenyl |
| | 535.2 inf. | 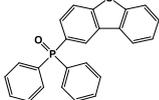 | 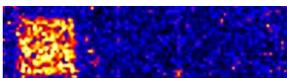 − O$_2$ |
| | 585.1 inf. | 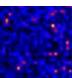 | 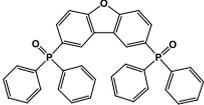 + O |
| | 645.2 inf. | 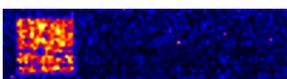 | 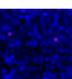 + Phenyl |
| | 1528.5 2.1 | 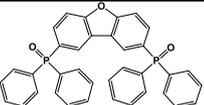 | 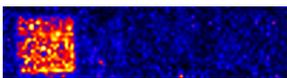 |
| | 1607.6 3.5 | 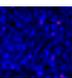 | 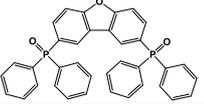 |
| | 1612.5 inf. | 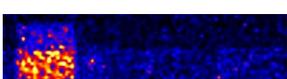 | |
| | 1624.6 inf. | 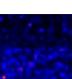 | 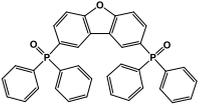 + OH |

*A/F fold : Mass intensity ratio between L10 aged pixel and fresh pixel. inf. = infinity, absent in fresh pixel.